# Long 48-cm ULE cavities in vertical and horizontal orientations for Sr optical clock

N. O. Zhadnov,[1,a)] K. S. Kudeyarov,[1] D. S. Kryuchkov,[1] G. A. Vishnyakova,[1] K. Yu. Khabarova,[1] and N. N. Kolachevsky[1,2]

AFFILATIONS

[1]P. N. Lebedev Physical Institute, Russian Academy of Sciences, Leninsky prosp. 53, 119991 Moscow, Russia

[2]Russian Quantum Center, Bol'shoi bul'var 30, stroenie 1, Skolkovo, 121205 Moscow, Russia

a)Author to whom correspondence should be addressed: nik.zhadnov@yandex.ru

ABSTRACT

*The development of an optical clock with ultimate accuracy and stability requires lasers with very narrow linewidth. We present two ultrastable laser systems based on 48 cm long Fabry-Perot cavities made of ULE glass in horizontal and vertical configurations operating at 698 nm. Fractional frequency instability of the beat signal between two lasers reaches $1.6 \cdot 10^{-15}$ at the averaging time of 1 s. We experimentally characterized the contribution of the different noise sources (power fluctuations, residual amplitude modulation, the Doppler noise, sensitivity to the shock impact) and found that in our case the laser frequency instability to a large extent is determined by an optoelectronic feedback loop. Although the vertical configuration proved to be easier to manufacture and to transport, it turned out that it is much more sensitive to acoustics and horizontal accelerations compared to the horizontal one. Both laser systems were transported over a 60 km distance from the Lebedev Physical Institute to the Russian National Metrology Institute (VNIIFTRI) where they serve as local oscillators for spectroscopy of the clock transition in the recently developed strontium optical clock.*

## 1. Introduction

Coherence, phase stability, and narrow spectral line are the key properties of any interrogation laser source for precision optical measurements. Laser systems with the fractional frequency instability at 1 second at the level of $10^{-15}$ and lower are widely used for precision spectroscopy[1], many-body physics[2], laser interferometry[3], astronomical spectroscopy[4], measurements of fundamental constants variation[5], and space missions to measure the geopotential[6].

Moreover, lasers with ultranarrow spectral linewidth are the unique instruments which allow precision spectroscopy of strongly forbidden atomic transitions in optical clocks[7–9]. Due to the periodicity of interrogation, high-frequency noise of the laser is aliased down to lower frequencies, limiting the clock stability at long averaging times. This phenomenon, known as Dick effect[10], poses tight limits on the high-frequency phase noise of the laser. Progress of modern optical atomic and ion clocks is significantly stimulated by developments in the field of ultrastable lasers. Note, that the development of the state-of-the-art ultrastable laser system in the early

2000s[11] paved a way for practical realization of the first passive $^{199}$Hg$^+$ single-ion optical frequency standard[12]. Today the most stable laser system based on a silicon cryogenic reference cavity[13] enables high performance of the modern Sr lattice clock at low $10^{-18}$ level with the averaging time of 1000 s[8].

Any laser frequency stabilization system requires a frequency discriminator of the highest possible Q factor. The most widely used frequency discriminator is a high-finesse monolithic evacuated Fabry-Perot (FP) cavity[13–15]. The laser frequency can be locked to a cavity eigenmode by the various methods, including the most widespread - Pound-Drever-Hall (PDH) technique[16]. Stability of the cavity resonant frequency relies on the special mechanical properties of its constituent materials, rigorous thermal stabilization, isolation from environmental noise, and is fundamentally limited by the thermal motion of cavity[17,18]. Among the alternatives to FP cavities, there are whispering-gallery mode microcavities[19], spectral hole burning[20], Brillouin lasers with fiber cavities[21] and locking to narrow transitions in thermal atomic beams[22].

One of the intriguing applications of ultrastable lasers is the search for the dark matter. According to some dark matter theories, the dark matter consists of scalar fields formed by ultralight particles ($mc^2 \ll 1\ eV$). Such fields lead to harmonic oscillations of fundamental constants, causing oscillations of material objects dimensions, and therefore, perturbing resonant frequencies of optical cavities. An experiment with an ultrastable cavity and an imbalanced Mach-Zehnder interferometer have been already implemented for the search of dark matter in the range of $4.1\cdot10^{-11} \div 8\cdot10^{-10}$ eV[23]. Another proposal is based on the idea of using a pair of FP cavities - a monolithic one and one with freely hanging mirrors - to detect the dark matter in the range $10^{-13} \div 10^{-11}$ eV[24]. Relying on different sensitivity of atomic transition frequency and cavity resonance to the presence of the dark matter one can use optical clock[25] to examine the range of masses $10^{-16} \div 10^{-21}$ eV.

In this article, we describe two ultrastable lasers operating at the wavelength of 698 nm actively stabilized to two long FP cavities of different configurations operating at the room temperature. The laser systems were developed at the P. N. Lebedev Physical Institute (Moscow, Russia) and then transported 60 kilometers away to the Russian National Metrological Institute VNIIFTRI where they are implemented as interrogation lasers in the new generation of Sr lattice clock systems (the description of the previous version of VNIIFTRI Sr clock can be found in [26]). We perform detailed characterization of laser systems and compare their characteristics.

## 2. Laser systems setup

An ultrastable laser system for an optical clock consists of a laser source, reference high-finesse Fabry-Perot cavity placed in a vacuum chamber, and an optoelectronic feedback loop that locks the laser frequency to the cavity mode. In this chapter, we describe all the constituent parts of the systems which were developed. Distinctive features of our systems are the 48-cm long cavities made of ULE glass in horizontal and vertical configurations. The horizontal FP cavity of such length was reported earlier[14], but, to the best of our knowledge, the vertical cavities of such length have not been studied up to the date.

## 2.1. Design of the cavities

For the fabrication of cavity spacers, we used ULE glass (Corning) - the traditional material for designing ultrastable systems due to its room-temperature zero thermal expansion point. We selected the spacer length of 480 mm which corresponds to the FP free spectral range (FSR) of 312.5 MHz. Increasing the length of the cavity reduces the thermal noise limit according to the expression[27]:

$$\sigma_y \sim \frac{T^{1/2} \varphi_{coat}^{1/2}}{L^{5/4} \lambda^{1/2} E_{sub}^{1/2}}, \quad (1)$$

where $T$ is the working temperature, $\varphi_{coat}$ – the loss angle of dielectric multilayer coatings, $L$ – the length of the cavity, $\lambda$ – the wavelength, $E_{sub}$ – the Young's modulus of the substrate. This expression is true when the Brownian thermal noise of the mirror coatings dominates over the other sources of thermal noise, which is the case for substrates made of high-Q fused silica. Increasing the length from 78 mm (our previous cavity generation[28]) to 480 mm allowed to decrease their thermal noise limit from $10^{-15}$ to $10^{-16}$ according to (1). The increase of the length reduces the fractional contribution of the mirror position fluctuations and makes the mode spots on the mirrors larger, so thermal fluctuations are averaged more effectively[18].

We decided to implement two different cavity configurations – the horizontal and the vertical – for several reasons. The horizontal cavity configuration with specially adjusted suspension points (see further for details) has been already implemented by several groups worldwide[14,29] and has proven its ultimate characteristics. Among disadvantages, one can enumerate certain manufacturing difficulties (ULE spacer shaping and drilling, bulky suspension, rectangular-shaped vacuum chamber). On the other hand, our group has a positive experience with vertical mid-plane suspended cavities of cylindrical symmetry[28,30], which are significantly more friendly for manufacturing and show a similar response to vertical accelerations[18]. Due to technological reasons (drilling) and the material price we also decided to assemble the vertical cavity from three separately manufactured parts.

The vertical cavity consists of three parts: one cylinder (the central part) and two truncated cones (the ends), bonded to each other by an optical contact. The cavity has a traditional vertical mount: three support points are placed symmetrically in the center-of-mass plane (fig. 1). Such a mount suppresses influence of the vertical vibrations, while sensitivity to the horizontal vibrations is reduced due to the bicone shape. During the fabrication process, the most crucial task was to make two cones of ULE glass of exactly equal shape, which is important to achieve the lowest vibrational sensitivity in the vertical direction. A certain challenge was also the large-area optical contact which required accurate adjustment of corresponding optical surfaces. On the other hand, drilling of the central hole of 12 mm in diameter and 15 cm in depth did not pose any technical difficulty.

The horizontal cavity has a shape of rectangular parallelepiped and is manufactured of a whole piece of ULE glass (fig. 2). The whole solid ULE glass block of 50-cm in length costs significantly higher than three separate blocks (15 cm, 15 cm and 20 cm in length). The most challenging part of fabrication was drilling the central hole with the diameter of 12 mm in the 48-cm cavity spacer. Drilling was carried out at the Lebedev Institute with the diamond holesaw half the length of the cavity deep from two sides. In order not to damage the big piece of ULE glass the drilling was done slowly and took around 30 hours in the whole. An important trick to drilling deep and straight holes in ULE glass is to apply the hammer drill of a certain strength to the holesaw. Otherwise, efficient drilling stops after approximately 1 cm of hole depth, and one has to apply extreme pressure on the holesaw which bends it. The mismatch between two opposite holes at the center of the ULE block was about 0.1 mm.

The horizontal cavity is supported at four points by a frame made of invar – a nickel-iron alloy with CTE around $10^{-6}$ 1/K. A low-thermal expansion support is needed to neutralize the CTE difference between ULE and aluminum of the vacuum chamber walls. To make the support softer and less transparent to acoustic waves, viton balls were placed at the support points between the spacer and the frame. The other important function of the balls is to equalize reaction forces at the four support points, which can differ due to imperfections of machining and temperature gradients in the vacuum chamber. This effect leads to uneven deformation of the cavity body and can be flattened by soft rubber supports. The viton balls were also put between the invar frame and the supporting pillars. There is no rigid fixation of the horizontal cavity.

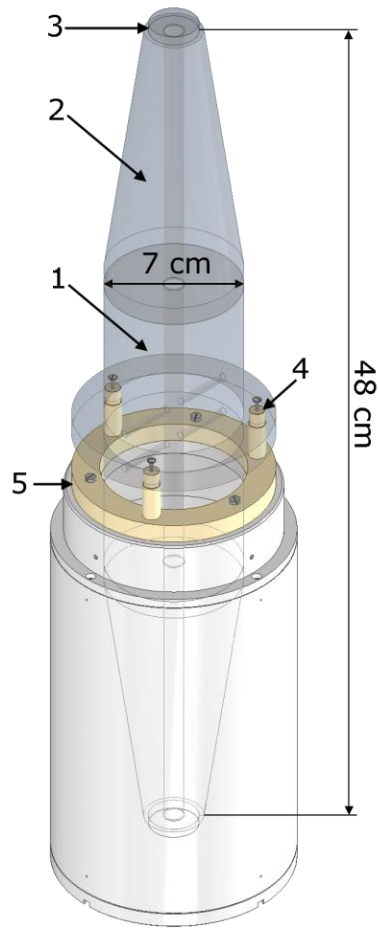

*Figure 1.* The design of the vertical cavity and its support system. 1 – central part of the cavity, 2 – one of two end parts of the cavity, 3 – upper mirror of the cavity, 4 – one of three mid-plane support points, 5 – support system made of PEEK. Three parts of the vertical cavity are bonded to each other via optical contact.

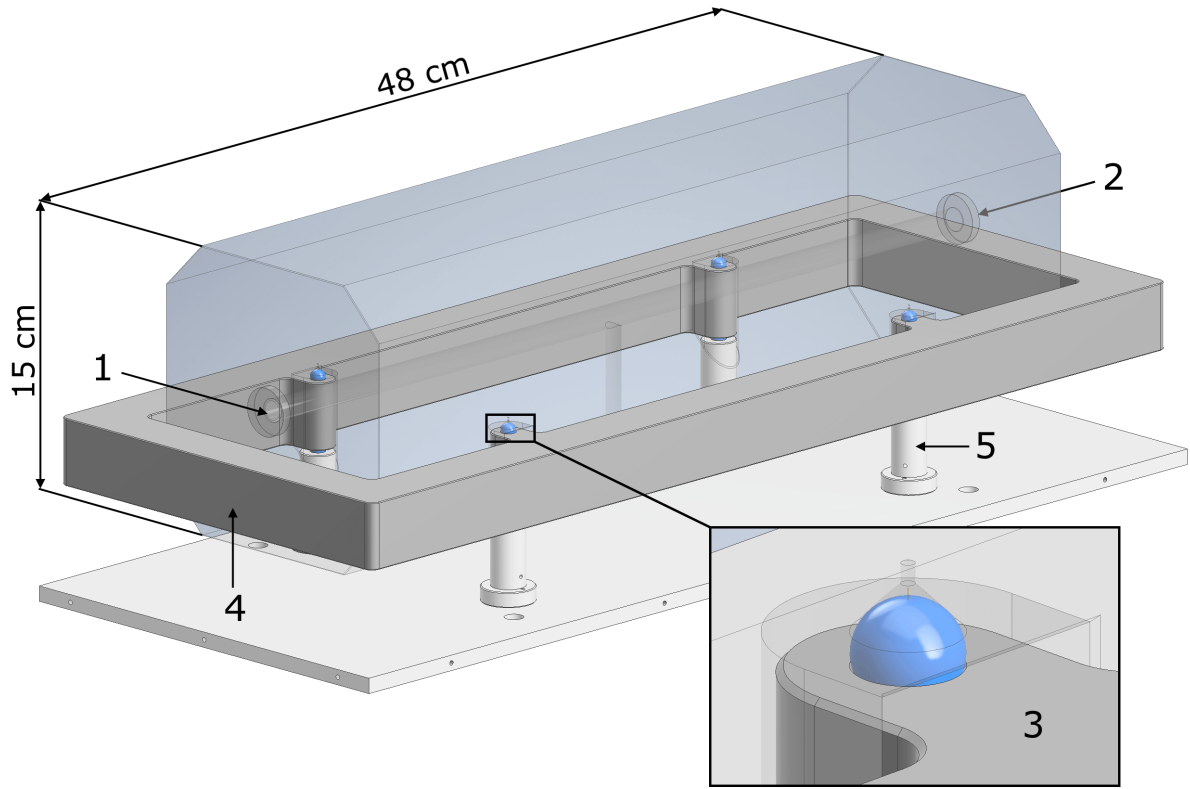

*Figure 2.* The design of the horizontal cavity and its support system. 1, 2 – the cavity mirrors, 3 – the support point with the viton ball, 4 – the invar support frame, 5 – the aluminum pillar. The viton balls are put between the cavity and the frame and between the frame and the pillars.

We performed finite element modeling of the horizontal cavity to find optimal positions of its supports. It turned out that if the central point of the support surface is placed at 99.0 mm from the end of the cavity and at 59.9 mm from the bottom of the cavity, vertical external force does not cause any tilt or relative translation of the mirrors. According to our modeling, the vertical cavity possesses the largest vibrational sensitivity in the horizontal direction (170 kHz/gal) and has sensitivity in the vertical direction of less than 1 kHz/gal. Vibrational sensitivity of the horizontal cavity in all directions is less than 5 kHz/gal[18].

Mirror substrates for both cavities are made of fused silica. Using a material with a high mechanical quality factor instead of ULE substrates makes it possible to further reduce the thermal noise[31]. ULE glass used for cavities fabrication has zero CTE point higher than the room temperature. However, the difference in CTEs of ULE and fused silica led to a zero-point shift to lower temperatures due to temperature-dependent deformation of the mirror substrate at the place of the optical contact. If one needs to suppress this effect, it can be made by additional ULE rings[32], but in our case the effect was not large because of the spacer length. As a solution, we have set the temperature of the FP cavities lower by slightly modifying the temperature stabilization system.

For each cavity, we use a pair of one flat and one spherical mirror with the 500 mm radius of curvature. The highly reflective coating consists of a multilayer Bragg structure of $SiO_2/Ta_2O_5$. The mirrors and the coatings were manufactured by SigmaKoki Company (Japan). Our ringdown measurements showed that the finesse of the vertical FP cavity equals 120000 (the cavity linewidth equals 2.6 kHz). The finesse of the horizontal cavity equals 80000, which corresponds to the resonance linewidth of 3.9 kHz. Unfortunately, the finesse occurred to be much lower compared to world's records (approaching $10^6$ for the best cases[23]), we were strongly dependent on the manufacturer. Only recently we managed to demonstrate FP cavity finesse approaching 100000 using ion beam sputtering facilities directly at the Lebedev Institute. Lower FP cavity finesse reduces frequency discriminator Q-factor and, as a result, poses stringent requirements for the locking electronics.

### 2.2. Vacuum chambers

Vacuum chambers of both horizontal and vertical cavities are manufactured from passivated duralumin and contain four internal thermal shields (fig. 3). Outer walls, 2 cm thick, provide big thermal capacitance of the chamber and, taking into account high thermal conductivity of duralumin, make it weakly sensitive to temperature fluctuations in the lab. A large mass of the chambers also reduces sensitivity to low-frequency vibrations. Chambers parts were covered with chromate conversion coating to passivate the surface. The chambers and the internal thermal shields are provided by anti-reflection coated optical ports, slightly tilted from the optical axis to prevent etalon effects. All vacuum sealings are made of indium wire.

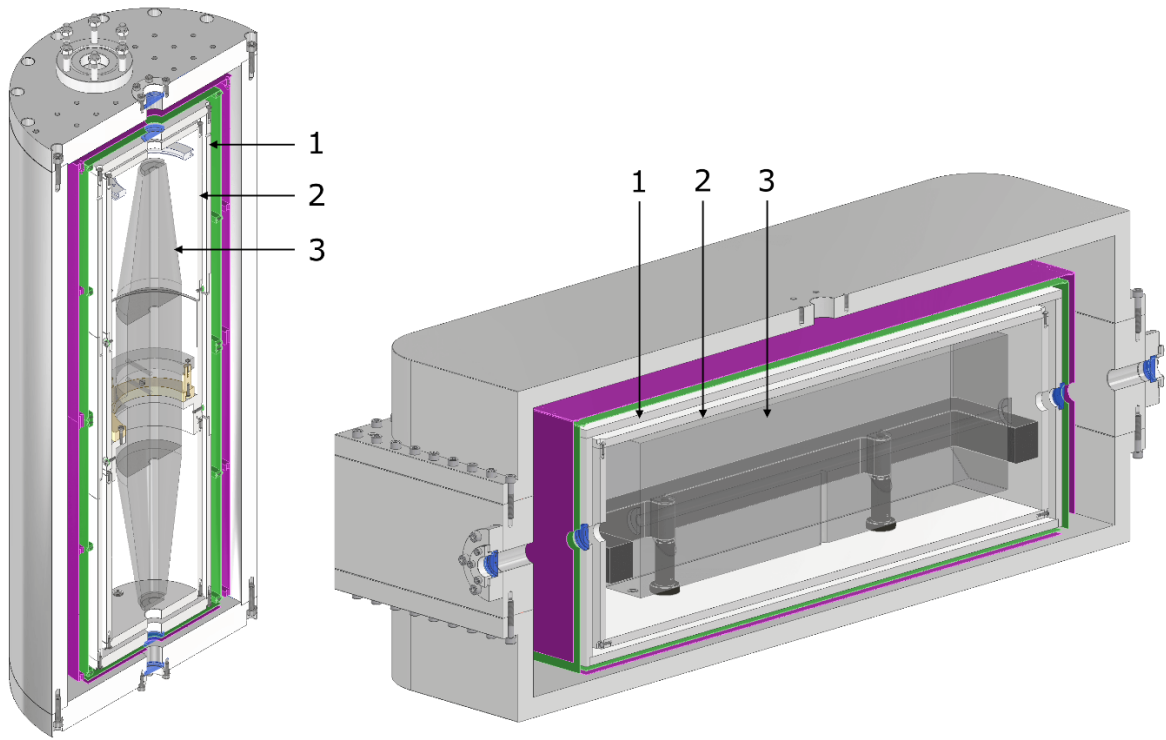

*Figure 3.* The vertical (left) and the horizontal (right) vacuum chambers. 1 – the temperature-stabilized shield, 2 – the thermal shield which serves as a thermal low-pass filter, 3 – the cavity. Two other thermal shields (highlighted in green and purple) are manufactured from thin duralumin sheets.

Vacuum chambers were pre-evacuated with a turbomolecular pump and annealed at the temperature 100°C. After annealing vacuum was maintained by the ion-getter pump NexTorr D 100-5 with the pumping speed of around 100 l/s. In the operating regime, pressure in the vertical system equals $1.4 \cdot 10^{-8}$ mbar, and in the horizontal system - $3.4 \cdot 10^{-8}$ mbar. This level is enough to neglect the influence of the refraction index variation of residual gas.

Temperature stabilization of the cavities is done by controlling the temperature of the thermal shield 2 by a pair of double stage Peltier elements (TEC). By monitoring the frequency of cavity transmission peak, we determined the zero CTE point for each of the cavity. The zero CTE points of vertical and horizontal cavities turned out to be at 17.2°C and 15.1°C, correspondingly. In this regime, the power consumption of the TEC equals approximately 5 watts at the room temperature of 21°C. Heat transfer from the thermal shield 2 is provided by copper buses glued by Torrseal glue (fig. 4). The temperature of shield 1 is measured by the sensor AD590 and controlled by a PI controller with a precision of better than 10 mK. We do not expect significant temperature gradients on the cavity body caused by non-symmetric cooling. First, the zero CTE temperatures are not too far from the room temperature, and additional thermal shield 2 plays the role of a temperature low-pass filter. Our previous experience with one-point heat transfer for the temperature stabilization of 78-mm length cavities[28,30] was successful.

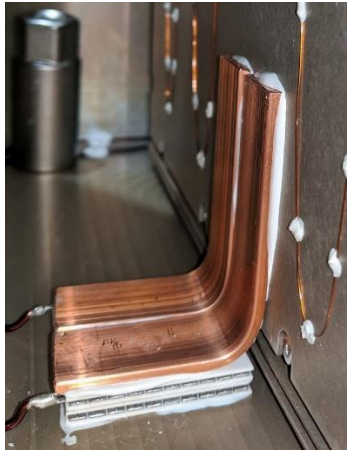

*Figure 4* The double-stage TEC package cools the internal part of the vacuum chamber. Heat is transferred via copper buses attached to the thermal shield 2 with vacuum glue.

After assembling, the vacuum chamber of the vertical system is placed at the center of an optical breadboard where the laser and all optical elements are placed (fig. 5a). Optical breadboard of the horizontal system is placed directly on top of the vacuum chamber (fig. 5b). Mechanical stability of the chambers is provided by the active vibration isolation systems Table Stable TS-300. It operates in the frequency range from 0.7 Hz to 300 Hz and provides at frequencies higher than 10 Hz the transmissibility of -40dB.

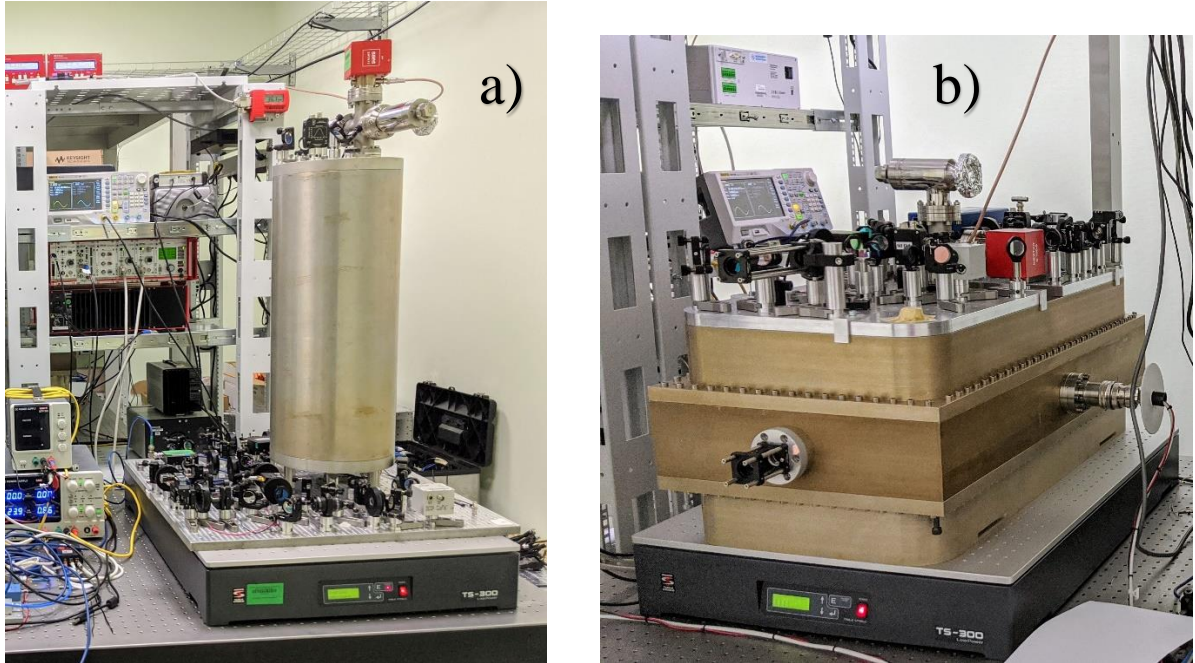

*Figure 5.* The vertical (a) and the horizontal (b) systems assembled in the laboratory.

### 2.3. Laser stabilization systems

The laser frequency is locked to the cavity mode using the PDH technique. Optoelectronic feedback loops of both laser systems are generally the same. The systems include 698 nm Toptica DL pro diode laser with an external cavity. The laser frequency can be tuned by changing injection current, the temperature of the diode, or voltage on piezo that moves the diffraction grating. The laser emission goes through an acousto-optic modulator (AOM), which also serves as an optical isolator to prevent etalon effects and can be used to stabilize optical power. Phase modulation of laser radiation at the frequency of around 20 MHz is provided by a free-space electro-optic modulator based on birefringent lithium niobate crystal. The rear facet of the crystal is wedged which provides the spatial separation of ordinary and extraordinary beams. It suppresses the residual amplitude modulation (RAM)[33,34] that can impact the frequency stability. A typical light power delivered to the cavity is 20 µW. The cavity transmission is about 2% for each of the systems which also indicates certain manufacturing problems during coating deposition. Laser light reflected by the cavity is detected by an avalanche Si photodetector. Servo signal is generated by a fast analog PID-controller Toptica FALC and fed to laser diode current control. Feedback loop bandwidth is characterized to be 460 kHz for the vertical system and 540 kHz for the horizontal system.

### 3. Characterization of laser systems

Here relative frequency stability of lasers and the influence of different noise sources is discussed. We measured the beat signal of two laser systems locked to corresponding ULE cavities (vertical, horizontal). During all measurements, laser systems were placed on different optical tables, at 5 m from each other which allowed them to be considered as independent. The beat note unit was assembled on the optical breadboard of the vertical cavity. The laser light of the horizontal system was transmitted to the vertical one via stabilized fiber link with the compensation of fiber-induced phase noise[35].

### 3.1 Relative frequency drift, spectral linewidth, and relative frequency instability of the vertical and the horizontal systems

The frequency drift of the ULE glass cavities is associated with the compression of the cavity spacer, usually explained by glass degassing. The relative frequency drift of the two systems was studied in [36] and turned out to be about 200 mHz/s. We expect that the drift will be reduced during aging of the systems and will reach tens of mHz/s in a few years. The main focus of the current work was to study the stability at averaging times 0.1 – 10 seconds which is important for interrogation of narrow clock transitions.

The power spectrum of the beat signal corresponding to the measurement time of 1 s is presented on fig. 6. The FWHM value of Lorentzian fit is 1.47 Hz, which is slightly larger than expected for the thermal-noise limited FP cavities.

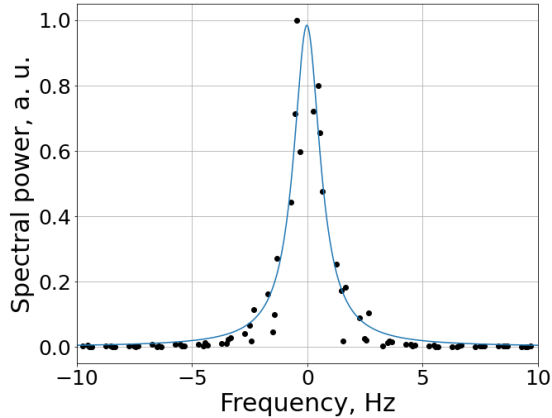

*Figure 6.* The power spectrum of the beat signal between two laser systems locked to the vertical and the horizontal cavities. The beat signal was recorded with an oscilloscope four times (with the recording length of 1 s), each dataset was Fourier transformed with the Hann window. The center of each line was determined and shifted to the origin. The resulting data (black dots) is fitted by the Lorentzian with FWHM of 1.47 Hz (blue line).

The relative frequency stability of the systems was studied by recording the beat signal frequency using the FXE frequency counter (by K+K company) with no dead time in the Λ-mode. Typical frequency track with 5 ms gate time and linear drift removed is shown in fig. 7. The corresponding modified Allan deviation (MDEV)[37] of the fractional beat frequency and the power spectral density (PSD) of phase noise is shown in fig. 8. Frequency fluctuations are averaged down over short times, for the interval 40 ms - 3 s MDEV reaches a flicker noise floor at $1.6 - 2 \cdot 10^{-15}$. At longer times the instability increases, which points to the prevalence of the random frequency walk. It is worth noting that we provide values for instability of the beat frequency, so individual instability of each laser system is better by a certain factor. Assuming that the systems are equivalent, the factor equals $\sqrt{2}$, but it can happen that most of the instability is due to one of the systems. Sharp peaks on PSD in the range 20-100 Hz can be attributed to acoustics.

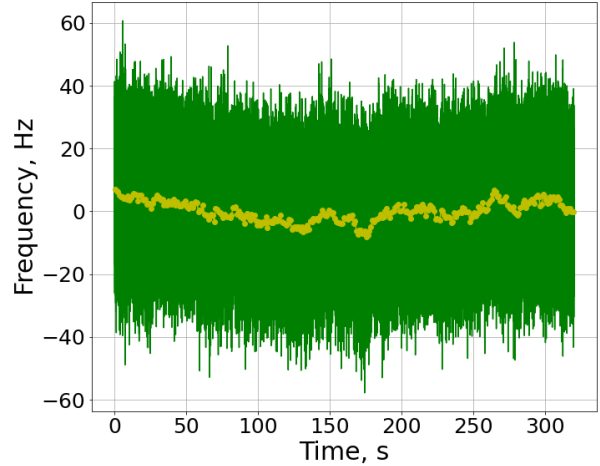

*Figure 7.* The frequency reading of the beat signal between the vertical and the horizontal laser systems operating at 698 nm (linear drift is subtracted). Green – data from the FXE frequency counter in the $\Lambda$-mode with the gate time 0.005 s, yellow – the data averaged for 1 s.

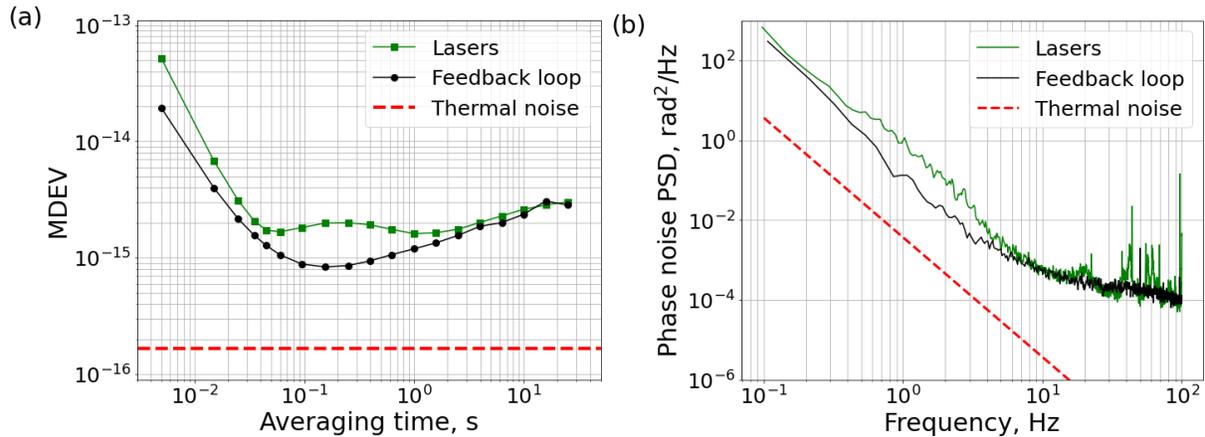

*Figure 8.* The modified Allan deviation (a) and the phase noise power spectral density (b) of the beat signal between two 698 nm stabilized laser systems. Red curves – two lasers are individually locked to the vertical and the horizontal cavity, MDEV corresponds to the dedrifted frequency reading from fig. 7. The black curves - both lasers are locked to the vertical cavity for characterization of the optoelectronic feedback loop (section 3.2). The red dashed curves – the expected thermal noise level.

### 3.2 Feedback loop characterization

Optoelectronic feedback loop contributes to the frequency instability of an actively stabilized laser via electronic noise (drifts of the offsets, amplification, etc.) and optical imperfections (optical etalons, power fluctuations, etc.). To characterize the noise of the optoelectronic feedback loops we performed the comparison between two individual lasers locked to the single (vertical) cavity, similar to the method described in [36]. We locked two lasers to two frequency-separated TEM00 modes of the vertical cavity spaced by two FSRs. In this experiment lasers were actively frequency stabilized using its individual PDH optoelectronic scheme. In this configuration, fluctuations associated with the instability of the cavity become common to both laser systems and cancel out by heterodyning. In order to decouple feedback loops from each other, polarizations of optical beams coupled to the cavity were perpendicular, while PDH modulation frequencies were significantly detuned from each other. Even with these precautions we still observed some crosstalk between two feedback signals if the laser beams were coupled from one side of the cavity. After readjusting the scheme such that two laser fields were coupled from the opposite sides of the cavity, the crosstalk was completely suppressed.

Results of the measurement are presented in the fig. 8. One can see that for the averaging time periods up to 600 ms, the optoelectronic noises contribute to approximately half of the net instability. The corresponding MDEV reaches the minimum of $8.4·10^{-16}$ at 155 ms and then starts growing again. Excluding the impact of fluctuations associated with the cavities themselves removes the bump in MDEV observed at the averaging time interval 100-500 ms. It indicates that cavity length fluctuations dominantly contribute at this time interval. Note, that for the longer averaging times (> 2 s) the optoelectronic fluctuations seem to play the dominant role in the frequency instability.

It occurred somehow disappointing, but our efforts to localize the source of these fluctuations and to improve the system characteristics were not successful. Of course, relatively low finesse of our cavities (approximately 3 times lower compared to PTB ones[14]) impose additional requirements to the quality of electronics. Indeed, decreasing the cavity finesse and coupling efficiency η makes the slope of the error signal flatter according to the expression

$$D \sim \frac{\sqrt{\eta}}{\Delta v},$$

where $D$ is the slope of the error signal and $\Delta v$ is the reference cavity linewidth[39]. To solve this problem one should set up elaborated home-made electronics and replace the high-finesse cavity mirrors, which we plan to implement in the next steps of the project.

### 3.3 Residual amplitude modulation

Phase modulation, necessary for the PDH method, leads to the residual amplitude modulation of the optical signal at the modulation frequency. Fluctuations of RAM depth and phase, that lead to instability of the error signal offset[33], can be caused by polarization fluctuations and parasitic etalons. Separation of ordinary and extraordinary beams in the EOM crystal leads to significant suppression of RAM and its impact on the laser frequency stability. Still, we decided to test two different methods of phase modulation: using the EOM and using direct modulation of the laser diode current. The current modulation caused ten times larger RAM depth while the amplitude of the error signal remained the same. The laser frequency stability measurements showed that there was no difference between the two cases. Making a reasonable assumption that RAM fluctuations should be roughly proportional to its amplitude, we conclude that RAM fluctuations do not have a noticeable contribution to the observed instability. Nevertheless, further research is needed to accurately determine the contribution of RAM to the frequency instability of the laser systems.

### 3.4 Frequency sensitivity to optical power circulating in the cavity

One of the possible sources of frequency instability in ultrastable lasers is a variation of light power circulating in the cavity caused by e.g. a change of the coupling efficiency or the laser intensity. Absorption of light by mirror coatings and substrates leads to heating, thermal deformation, and therefore, to the shift of the FP cavity frequency.

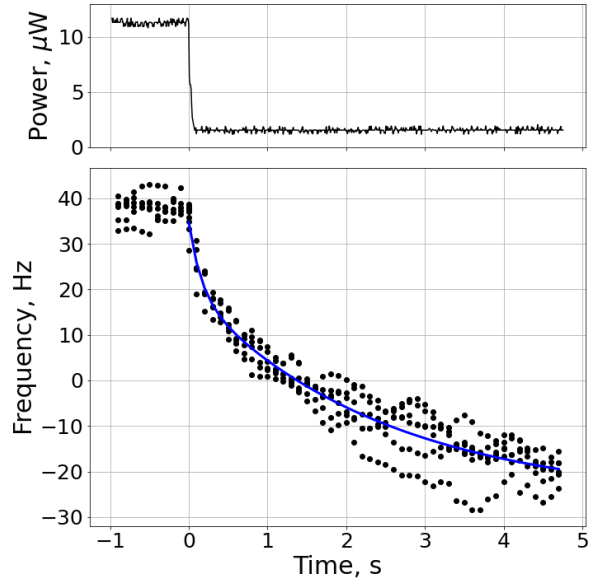

*Figure 9.* Frequency response of the vertical cavity to instantaneous change of the incoupling power. Top: change of the transmitted power (from 11.5 µW to 1.6 µW). Bottom: the frequency response (linear drift subtracted). Black dots - data of six consecutive measurements, blue line – bi-exponential fit. As the frequency reference the horizontal cavity is used.

To characterize the dependency of the cavity mode frequency on the coupled laser power, the following procedure was performed. We continuously recorded the beat signal frequency between two independent laser systems (the horizontal, the vertical). At a certain moment, the power coupled to one of the cavities was changed abruptly with the help of the AOM drive. It is important for this experiment to accurately adjust the feedback loop set point and offsets such that the change of the error signal amplitude (proportional to the light power on the PDH photodiode) will not cause significant set point shift.

The observed frequency response of the cavity can be fitted with the bi-exponential function $A_1 e^{\frac{-t}{\tau_1}} + A_2 e^{\frac{-t}{\tau_2}}$. It was observed previously that the fast exponent corresponds to the thermal response of the mirror coatings, while the slow exponent – to thermal effects in the substrates[40]. Typical frequency response of the vertical cavity to the power change is presented in fig. 9. Fitting the whole accumulated dataset, we obtained the following sensitivity for the vertical cavity: $A_1 = 1.5$ Hz per 1 µW of transmitted power for exponent with $\tau_1 = 0.2$ s and $A_2 = 5$ Hz/µW for $\tau_2 = 2$ s. The horizontal cavity demonstrates very similar behavior.

Previously reported 48-cm long horizontal cavity at PTB possesses the frequency sensitivity to the transmitted laser power of 120 Hz/µW[14]. It is 25 times higher than in our case. Although it is difficult to compare PTB and LPI systems (the outcoupler transmission was not published), we attribute lower sensitivity in our case to 3.5 times lower finesse and higher absorption in the mirrors.

Our results show that even without active stabilization of the optical power coupled to the cavity, the frequency instability induced by fluctuations of the optical power is below $10^{-16}$ at the averaging times of 0.01 s - 100 s. The active power stabilization using an AOM makes this effect completely negligible in our configuration.

### 3.5 Cavity Doppler noise

Vibrations can influence the frequency of the cavity mode not only by a change in the distance between the cavity mirrors caused by the deformation of the spacer, but also by the movement of the cavity as a whole due to the Doppler effect. This effect can be measured by an interferometric scheme, similar to the used for fiber noise compensation[35], and can be cancelled by a feedback loop using. The measured fractional frequency instability caused by the Doppler effect is at the level $10^{-15}$ at 100 ms (fig. 11) and averages down to lower numbers. Active compensation of this effect did not improve the frequency stability of our laser systems.

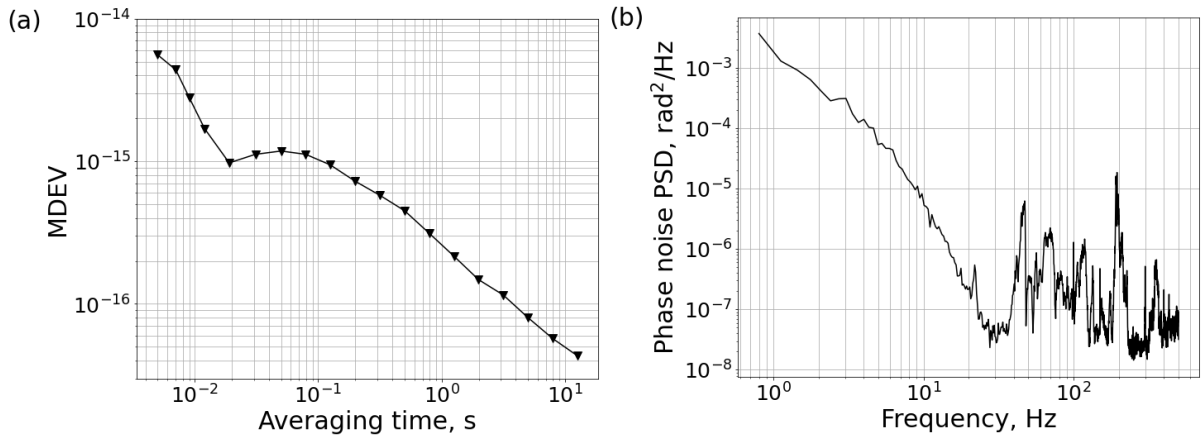

*Figure 10.* Modified Allan deviation (a) and phase noise power spectral density (b) of the Doppler noise induced by movement of the cavity as a whole.

### 3.6 Response to the shock impact

Vibrational sensitivity is one of the key limiting factors restricting the design of the cavity: the longer the cavity spacer, the higher vibrational sensitivity it possesses. To test the reaction of our systems to vibrations, we performed a simple impulse response experiment. One of the systems was underwent a shock action of a small rubber ball, free-falling on the optical breadboard while the beat frequency between two laser systems was continuously recorded. Elastic collision with a ball induces a broad spectrum of vibrations, which impacts the beat frequency via changes of the optical mode frequency and the Doppler effect.

In the first experiment the ball of 2 cm in diameter and was first dropped from the 15 cm height on the optical breadboard of the vertical system, in the second - on the breadboard of the horizontal one. The beat frequency shows that the vertical cavity support structure possesses the frequency resonance at 62 Hz, and the horizontal one – at 387 Hz (fig. 10). The measured Q-factors of these resonances are $Q_{vert} = 45$ and $Q_{hor} = 55$ correspondingly. Although the amplitude of the response was larger for the vertical cavity, it does not allow to quantitatively compare vibrational sensitivities of two cavities. Nonetheless, our experience with cavities of different orientation indicates that the vertical configuration possesses a significantly higher response to the acoustic noise in the laboratory, than the horizontal one. These observations are in agreement with the results of finite element modelling of cavity deformations. To isolate the laser systems from acoustic noise, sealed boxes made of thick plywood were constructed.

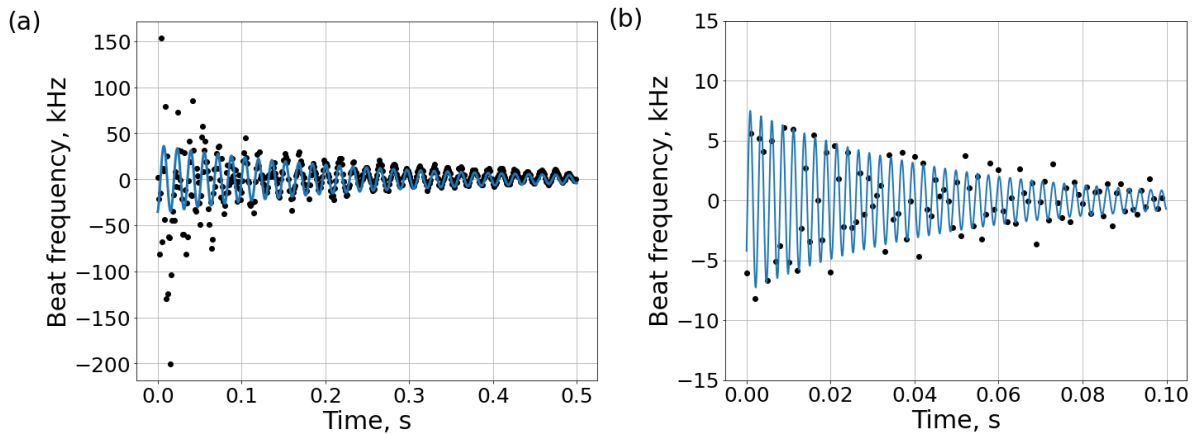

*Figure 11.* Vibrational response of stabilized lasers beat frequency under the action of a falling rubber ball (2 cm in diameter, from 15 cm height) on the corresponding optical breadboard: (a) – for the vertical system, (b) – for the horizontal system. Black dots – experimental frequency response, blue – fit by the exponentially damped sinusoidal function.

Direct measurement of the vibrational response function is a complicated task because of the difficulty to directly measure the cavity acceleration. These measurements have been performed previously for a spherical cavity only, which allowed implement an active feedback loop to compensate vibrations[41].

## 4. Discussion and outlook

In this work we present two ultrastable laser systems based on long 48 cm room-temperature ULE cavities operating at 698 nm for Sr optical clock, oriented vertically and horizontally. The fractional frequency instability of each individual system is lower than $2 \cdot 10^{-15}$ at the averaging times 40 ms – 3 s, which demonstrates good performance competitive to the short cavities previously studied in [28,30]. Still, it is approximately an order of magnitude higher compared to the fundamental thermal noise limit of $1.2 \cdot 10^{-16}$. One of the reasons is the excessive contribution of noise introduced by the optoelectronic systems which we studied experimentally. The corresponding contribution to the Allan deviation makes up a significant part on short time intervals, which equals to $8.4 \cdot 10^{-16}$ at the minimum and dominates on times > 2 s. We attribute it to the insufficient quality of the cavity high-reflecting mirrors providing the finesse of only about 100000. Together with relatively low cavity transmission compared to the best world's results [14,42] it puts stringent requirements on the feedback loop. We studied other important characteristics of the cavities, demonstrated relatively low sensitivity to laser power fluctuations and qualitatively characterized the impact of shock vibrations to the systems. The observed hump in the Allan deviation plot for the time interval 0.5 s – 5 s is likely associated with the coupling of the vertical cavity to the acoustic environment of the laboratory.

Although the vertical cavity system was significantly easier to manufacture (composite spacer, easy drilling, axial symmetry) and to transport compared to the horizontal one, the vertical cavity occurred to be much more sensitive to horizontal accelerations and, correspondingly, to acoustics. From the point of ultimate stability, the horizontal configuration seems to be more preferable despite several manufacturing challenges.

Both optical systems were successfully transported over a 60 km distance from LPI to VNIIFTRI using a regular truck, minor adjustment was required to set them to operation at VNIIFTRI. Several experiments reported in this paper were performed directly at VNIIFTRI, $^{87}$Sr clock transition spectroscopy was successfully performed using the horizontal system.

To reach ultimate performance and approach the thermal noise limit we plan to replace high-reflecting mirrors with better ones and to upgrade some electronic components. It should reduce the influence of optoelectronic feedback loop noise on the net frequency stability of the laser by an order of magnitude.

This study was supported by the Russian Foundation for Basic Research, project no. 19-32-90207.